\documentclass[12pt]{article}
\usepackage{color}
\begin{document}
\begin{center}
{\bf CASIMIR FRICTION FORCE AND ENERGY DISSIPATION FOR MOVING HARMONIC OSCILLATORS}

\vspace{1cm}
Johan S. H{\o}ye\footnote{johan.hoye@ntnu.no}

\bigskip
Department of Physics, Norwegian University of Science and Technology, N-7491 Trondheim, Norway

\bigskip
Iver Brevik\footnote{iver.h.brevik@ntnu.no}

\bigskip
Department of Energy and Process Engineering, Norwegian University of Science and Technology, N-7491 Trondheim, Norway

\bigskip
\end{center}

\begin{abstract}
The Casimir friction problem for a pair of dielectric particles in relative motion  is analyzed, utilizing a microscopic model in which we start from statistical mechanics for harmonically oscillating particles at finite temperature moving  nonrelativistically with constant velocity.  The use of statistical mechanics in this context has in our opinion some definite advantages, in comparison with the more conventional quantum electrodynamic description of media that involves the use of a refractive index. The statistical-mechanical description is physical and direct,  and the  oscillator model, in spite of its simplicity,  is nevertheless able to elucidate  the essentials of the Casimir friction.  As is known, there are diverging opinions about this kind of friction in the literature. Our treatment elaborates upon, and extends, an earlier theory presented by  us back in 1992. There we found a finite friction force at any finite temperature, whereas at zero temperature the model led to a zero force. As an additional development in the present paper we evaluate the energy dissipation  making use of an exponential cutoff truncating the relative motion of the oscillators. For the dissipation we also establish a general expression that is not limited to the simple oscillator model.
 \end{abstract}


\bigskip

\section{Introduction}
 The Casimir effect - the original paper being
Ref.~\cite{casimir48} - has attracted a formidable increase of
interest in the later years. For instance, as shown by a graph in
the 2005 Lamoreaux paper \cite{lamoreaux05}, the number of
citations per year to the 1948 Casimir paper has increased from
about 10 to well above 100 in the period from 1980 to 2010.

The standard  Casimir setup is that of two
parallel metallic or dielectric plates at micron or sub-micron
separations,
 the attractive force between them being calculated or measured.
 Later years have  seen a considerable progress in the analysis of
Casimir forces also between bodies of a more general shape, under
static conditions. Some recent general treatments can be found in
Refs.~\cite{milton01,milton04,buhmann07,bordag09}.

 A special sub-class of Casimir phenomena is that
of {\it friction}. The magnitude of the Casimir friction force
between plane surfaces in relative parallel motion  has attracted
considerable interest recently. The problem as such is not new; it
was studied a long time ago by
 Teodorovich \cite{teodorovich78} and by Levitov \cite{levitov89}. Polarization
currents fluctuating in the two bodies interact via the
electromagnetic field, transferring photons between the slabs.
Because of the relative motion the photons are subject to Doppler
shifts, and one should expect there to be a resulting dissipation of energy, meaning a
friction force. The works mentioned were based upon  macroscopic
theory of the electromagnetic field in a medium, involving use of a refractive index.
 Most of the more recent literature has been formulated along the same macroscopic lines.
  Some of the later papers can be found in
   Refs.~\cite{pendry97,pendry98,volokitin07,volokitin08,philbin09,philbin09a,dedkov08,dedkov08a,dedkov10,pendry10}.

A complicating factor in the macroscopic theory
is that there exists no natural rest inertial frame wherein the
system of bodies is collectively at rest. This may be a major
reason why diverging results are found in the current literature.
For instance, Philbin and Leonhardt find there to be no quantum
friction at all \cite{philbin09}, whereas most other papers find
the friction force to be nonvanishing.  Recent papers of Dedkov
and Kyasov \cite{dedkov10} and of Pendry \cite{pendry10} give
useful overviews and comparison with earlier results. The title of
Pendry's paper \cite{pendry10} is illustrative for this kind of
research: "Quantum friction - fact or fiction?".

Faced with this rather complex situation it becomes natural to
inquire to what extent light can be shed on the problem by
following an alternative  approach: Instead of starting from
quantum electrodynamics in a {\it medium } one can start with a
simpler microscopic model in which only moving {\it harmonic
oscillators} are involved. Such an investigation is the main theme
of the present paper. We shall consider two oscillators in
relative motion having a constant nonrelativistic velocity. They
are taken to represent a pair of polarizable particles.  As we
shall see, this oscillator model is in fact able to describe the
essentials of the Casimir friction force. Some years ago we
investigated such a microscopic model, for  instantaneous
\cite{hoye92}, as well as for non-instantaneous \cite{hoye93},
interactions. In view of the mentioned diverging
results obtained in this area of research,
 we find reason to consider this simple model anew, and to elaborate on it, the more so since results obtained in the model in essence should be
    valid for dielectric plates in
 relative motion also. The statistical mechanical approach implies that the Kubo formula \cite{kubo58,landau85} will play an important role. This way of
  approaching the Casimir friction problem is apparently not so well known in the Casimir community, but we find it
  right to emphasize the microscopic method's usefulness. It leads to physical results without use of a heavy mathematical
  formalism. The microscopic approach has been followed in a recent
  investigation by Barton also (personal communication).

We assume in the following thermal equilibrium conditions, and assume the interaction between the oscillators to be weak.
\vspace{0.5cm}

The results of the present paper can be summarized as follows:
\vspace{0.5cm}

$\bullet$ There is a finite Casimir force, at any finite temperature $T$.

$\bullet$ For the simple oscillator model  the Casimir force vanishes at $T=0$.

$\bullet$ The Casimir energy dissipation is calculated, identifying the dissipation with the work done. Assuming infinite motion (i.e., the same velocity for all times), it is not clear how to distinguish between reversible energy change and irreversible energy dissipation. We handle this problem by introducing a convergence factor  by means of which the interaction  is limited to a finite time interval. Therewith the energy dissipation is calculated unambiguously.

$\bullet$ The final expression for the energy dissipation, Eq.~(\ref{27}) below, has a validity beyond the limitations of the harmonic oscillator model. Since this is a new result, it would be of interest to have it verified also in other ways, if possible.

 We mention finally that techniques similar to those used in the present letter have been used, for instance, in studies of the dynamical Casimir force associated with longitudinal motion of the plates (thus different from the lateral motion considered here). More information can be found in review articles \cite{jaekel97,dalvit10}, with further references therein.

\section{Calculation of the friction force}

We will consider the quantum mechanical two-oscillator system
whose reference state is the one of uncoupled motion corresponding
to a Hamiltonian $H_0$. The equilibrium situation becomes
perturbed by a time dependent term which we will write in general
form as $-Aq(t)$, where $A$ is a time independent
operator and $q(t)$ a classical function of time
whose explicit form depends on the specific properties of the
system. The Hamiltonian becomes $H=H_0-Aq(t)$.
Moreover, we put
\begin{equation}
-Aq(t)=\psi ({\bf r}(t))x_1x_2, \label{1}
\end{equation}
where $\psi (\bf r)$ is the so-called coupling strength (i.e.
$\psi$ is the classical potential between the oscillators). The
separation between the oscillators is   $\bf r$, and $x_1,x_2$ are
the internal vibrational coordinates of the oscillators. When the
oscillators move with respect to each other the coupling has to
vary in time. With nonrelativistic constant relative velocity $\bf
v$ the interaction will vary as
\begin{equation}
-Aq(t)=[ \psi ({\bf r}_0)+{\bf \nabla}\psi({\bf
r}_0)\cdot {\bf v}t+...]x_1x_2, \label{2}
\end{equation}
when expanded around the initial position  ${\bf r=r}_0$ at $t=0$.
 The force between the oscillators, called $\bf B$,  is
\begin{equation}
{\bf B=-(\nabla}\psi ({\bf r}))x_1x_2. \label{3}
\end{equation}
We ought here to mention the following point. In a mathematical
sense
 the expansion (\ref{2}) requires $vt$ to be small.
Physically, we assume nevertheless Eq.~(\ref{2}) to hold for all
times, so that the interaction energy is taken to be proportional
 to $t$ for all values of $t$. The natural opportunity of choosing
 $q(t)=t$ in the
interaction (\ref{1}) thus has to be modified: as will be shown
below,  a convergence factor will be needed.

Another point worth noticing is that the expression (\ref{1})
corresponds to first quantization only. Quantum electrodynamic
processes such as emission and absorption of photons (second
quantization) are not accounted for by the present model. They
were considered, however,  in Ref.~\cite{hoye93}.

The equilibrium situation with both oscillators at rest is
represented by the first term in (\ref{2}). It gives rise to a
(reversible) equilibrium force. Thus the friction must be
connected with the second term. To simplify, we will for the
moment neglect the first term. By this the two oscillators will be
fully uncorrelated in their relative  position ${\bf r=r}_0$. The friction force,
 due to the time dependence of the interaction (\ref{2}), will
be a small perturbation upon the equilibrium situation. This
interaction leads to a response $\Delta \langle {\bf B}(t)\rangle$
in the thermal average of $\bf B$. And this is where the Kubo
formula, mentioned above, comes in
\cite{kubo58,landau85,brevik88}:
\begin{equation}
\Delta \langle {\bf B}(t)\rangle =\int_{-\infty}^t {\bf
{\phi}}_{BA}(t-t')q(t')dt' \label{4}
\end{equation}
(note that $\boldmath{\phi}_{BA}$ means a vector),
where the response function is given by
\begin{equation}
{\boldmath\phi}_{BA}(t)=\frac{1}{i\hbar}\rm{Tr} \{ \rho [A,{\bf
B}(t)]\}. \label{5}
\end{equation}
Here $\rho$ is the density matrix and ${\bf B}(t)$ is the
Heisenberg operator ${\bf B}(t)=e^{itH/\hbar}{\bf
B}e^{-itH/\hbar}$, where $\bf B$ like $A$ is time independent. Now
with (\ref{2}) and (\ref{3}), and with $q(t)=t$, expression
(\ref{5}) can be rewritten as
\begin{equation}
{\boldmath \phi}_{BA}(t)={\bf G}\phi(t), \label{6}
\end{equation}
with
\[ {\bf G}=({\bf \nabla}\psi )({\bf v\cdot \nabla}\psi), \]
\[ {\bf \phi}(t)=\rm{Tr}\{ \rho \,C(t)\}, \]
\begin{equation}
C(t)=\frac{1}{i\hbar}[x_1x_2,x_1(t)x_2(t)]. \label{7}
\end{equation}
Thus with Eq.~(\ref{4}) and $q(t')=t'$ the force
can be written as
\[ {\bf F}=\Delta \langle {\bf B}(t)\rangle ={\bf
G}\int_{-\infty}^t\phi (t-t')t'dt'={\bf F}_r+{\bf F}_f, \] where
\begin{equation}
{\bf F}_r={\bf G}t\int_0^\infty \phi(u)du \label{8}
\end{equation}
is part of the reversible force by which the part of the force that
represents friction is
\begin{equation}
{\bf F}_f=-{\bf G}\int_0^\infty \phi(u)udu. \label{9}
\end{equation}
Here the new variable $u=t-t'$ has been introduced. The ${\bf
F}_r$ can be interpreted  as a reversible force since it depends only upon position. This interpretation is consistent with the result obtained for the dissipation in Sec. 3 below; the ${\bf F}_r$ will not contribute to the net total dissipation.

If one again includes $\psi({\bf r}_0)$ one has
\[ {\bf G}t=({\bf \nabla}\psi)({\bf v\cdot \nabla}\psi)t \rightarrow
({\bf \nabla}\psi)[\psi({\bf r}_0)+t{\bf v}\cdot {\bf \nabla}
\psi +...]
\]
\begin{equation}
=({\bf \nabla}\psi)\psi ({\bf r}_0+ {\bf v}t), \label{10}
\end{equation}
where ${\bf r}={\bf r}_0+{\bf v}t$ is the position at time $t$. By
contrast, expression (\ref{9}) changes sign when the velocity
$\bf v$ changes sign, and it thus represents a friction force.
(Observe that the velocity in (\ref{10}) merely represents the
shift in position.) Equation (\ref{9}) is  the same as
result (2.11) in Ref.~\cite{hoye92} \footnote{There is a missing minus sign in Eq.~(2.11) and the ${\bf F}_r$ was not taken into account.}, and the Fourier transformed version of it and Eq.~(\ref{9}) above as well, is
\begin{equation}
{\bf F}_f=-i{\bf G}\frac{\partial \tilde{\phi}(\omega)}{\partial
\omega}\Big|_{\omega=0}, \label{11}
\end{equation}
where $\tilde{\phi}(\omega)=\int_0^\infty \phi(t)e^{-i\omega
t}dt~~$ (with $\phi(t)=0$ for $t<0$).

In Ref.~\cite{hoye92} the Fourier transformed version (\ref{11})
was used to obtain the explicit expression for the friction force.
Here, we instead will use  a different approach based on the  expression (\ref{9}). As in the
reference mentioned we then need the commutator (\ref{7}). This
entity again follows from the properties of quantized harmonic
oscillators. We introduce annihilation and creation operators $a$
and $a^\dagger$ with commutation relations $[a_i, a_i^\dagger]=1$
($i=1,2$; other commutators vanish). As usual, ${a_j} (t)=a_j
e^{-i\omega_jt}$ and ${{a_j}^\dagger(t)}={a_j}^\dagger
e^{i\omega_jt}$. With this the coordinates are
\begin{equation}
x_i=\left(\frac{\hbar}{2m_i\omega_i}\right)^{1/2}(a_i+a_i^\dagger)
\label{12}
\end{equation}
where $m_i$ and $\omega_i$ ($i=1, 2$) are the mass and
eigenfrequency of each oscillator. To obtain $\phi(t)$ from
(\ref{7}) we first have to calculate $\langle
n_1n_2|C(t)|n_1n_2\rangle$, where $|n_1n_2\rangle =
|n_1\rangle|n_2\rangle$ represents eigenstates with oscillators
excited to levels $n_1$ and $n_2$. One has, when
taking into account standard properties of the annihilation and
creation operators,
\begin{equation}
L_i \equiv \langle n_i|a_i^\dagger
a_i(t)+a_ia_i^\dagger(t) | n_i\rangle=(2n_i+1)\cos
(\omega_it)+i\sin (\omega_it), \label{13}
\end{equation}
from which the thermal average follows after some
computations as
\[ \phi(t)=\langle\langle n_1n_2|C(t)|n_1n_2\rangle\rangle   \]
\[= \frac{1}{i\hbar} \frac{\hbar}{2m_1 \omega_1}
\frac{\hbar}{2m_2 \omega_2} (L_1 L_2 -L_1^* L_2^* \]

\begin{equation}
= D \left[(2\langle n_1\rangle +1)\cos
(\omega_1t)\sin (\omega_2t)+(2\langle n_2\rangle+1)\cos
(\omega_2t)\sin (\omega_1t)\right]. \label{14}
\end{equation}
Here
\[ D=\frac{\hbar}{2m_1 m_2\omega_1\omega_2} .
\]
With energy levels $\varepsilon_n=(n+\frac{1}{2})\hbar \omega$ the
thermal average for the occupation numbers is
\begin{equation}
2\langle n_i\rangle +1=\coth (\frac{1}{2}\beta \hbar \omega).
\label{15}
\end{equation}
In Ref.~\cite{hoye92} expression (\ref{14}) was Fourier
transformed to obtain the friction force as given by (\ref{11}).
As an alternative we will here use expression (\ref{14}) directly
in Eq.~(\ref{9}). Then we get the integral
\[ \int_0^\infty te^{-\eta t}\cos(\omega_1t)\sin(\omega_2t)dt \]
\begin{equation}
= \frac{\eta \Omega_1}{(\eta^2+\Omega_1^2)^2}-\frac{\eta
\Omega_2}{(\eta^2+\Omega_2^2)^2} \rightarrow
-\frac{\pi}{2\Omega_2}\delta (\Omega_2), \quad \eta \rightarrow 0.
\label{16}
\end{equation}
Here $\Omega_1=\omega_1+\omega_2$ and
$\Omega_2=\omega_1-\omega_2$. As mentioned above a convergence
factor $e^{-\eta t}$ is needed, and the limit $\eta \rightarrow 0$
is taken. Then the $\Omega_2-$term  becomes a delta function with
prefactor determined by the  integral
\begin{equation}
\int_{-\infty}^\infty \frac{\eta
x^2}{(\eta^2+x^2)^2}dx=\frac{\pi}{2}. \label{17}
\end{equation}
From (\ref{14}) we also get this integral with $\omega_1$ and
$\omega_2$ interchanged. This will  then give the  result (\ref{16})
with opposite sign with respect to the $\Omega_2$-term. Adding up we get the difference of the
prefactors
\[ \coth(\frac{1}{2}\beta \hbar \omega_1)-\coth
(\frac{1}{2}\beta \hbar \omega_2)=-\frac{\sinh (\frac{1}{2}\beta
\hbar \Omega_2)}{\sinh(\frac{1}{2}\beta \hbar \omega_1)\sinh
(\frac{1}{2}\beta \hbar \omega_2)} \]
\begin{equation}
\rightarrow -\frac{\frac{1}{2}\beta \hbar
\Omega_2}{\sinh(\frac{1}{2}\beta \hbar
\omega_1)\sinh(\frac{1}{2}\beta \hbar \omega_2)}, \quad \eta
\rightarrow 0. \label{18}
\end{equation}
Multiplying (\ref{16}) with (\ref{18}) and including the factors $D$
and $\bf G$ the friction force becomes
\begin{equation}
{\bf F}_f=-\frac{\pi \beta \hbar^2({\bf \nabla}\psi)({\bf v\cdot
\nabla}\psi)}{8m_1m_2\omega_1^2\sinh^2(\frac{1}{2}\beta \hbar
\omega_1)}\delta(\omega_1-\omega_2), \label{19}
\end{equation}
which is also the result (3.14) of Ref.~\cite{hoye92}. Again one
notes that there is friction only when the oscillators have the
same frequency, and $\beta$ should be finite, i.e. $T>0$.

When $\beta \rightarrow \infty$, the expression (\ref{19}) vanishes.
 According to the present oscillator model there is thus no friction force at zero temperature.
 An objection against this result may be that it is somewhat singular due to the presence of the $\delta$-function. Thus its physical significance may not be obvious. However, $\eta$ can be kept finite. This will smooth out the $\delta$-function, and the $\Omega_1$-term in Eq.~(\ref{16}) will give a contribution too. Note that this will not change our conclusions about a finite friction force for $T>0$. (For finite $\eta$, i.e. interaction like a short pulse, there will also be a contribution for $T=0$ due to the $\Omega_1$-term in Eq.~(\ref{16}).) But here we will assume $\eta$ small by which the $T=0$ contribution can be disregarded.

 In Ref.~\cite{hoye92} the result (\ref{19}) for the friction force
was derived also by two other methods. These methods utilized the
path integral formalism of quantum systems at thermal equilibrium
\cite{hoye81}. The path integral can be identified with a
classical polymer problem where imaginary time is a fourth
dimension of length $\beta$. Thus the polymers stretch out in the
fourth dimension and form closed loops of periodicity $\beta$. For
harmonic oscillators the correlation function along the polymers
is obtained in a straightforward way.
With one of the methods the convolution of
the correlation functions of both oscillators were needed. The
resulting Fourier transform of this convolution was then identified
with the response function $\tilde{\phi}(\omega)$ used in
expression (\ref{11}) \cite{brevik88}.

With the other method full thermal equilibrium was utilized. Then
the relative motion of the oscillators was regarded as a harmonic
oscillator motion with low frequency $\omega_0\rightarrow 0$.
Again with the path integral one can obtain the Fourier transform
of the response function for the relative motion. The damping of
the relative motion, that can be related to this response
function, gives the friction force, and again the result
(\ref{19}) was recovered. Thus the three methods used in
Ref.~\cite{hoye92}, as well as the modification considered in the
 present paper, all lead to the same result, in contradiction to some other results in the literature, for instance that of Ref.~\cite{philbin09}.

The result for the friction force was also extended to the
situation with time-dependent or non-instantaneous interaction
\cite{hoye93}. Then the full thermal equilibrium method was
applicable to generalize the result. With the latter interaction
there was also a friction from the self-interaction of a moving
oscillator with itself.

\section{Dissipation of energy}
The presence of friction means dissipation of energy; the thermal energy of the system has to increase. However, this increase will be of second order in the velocity. As the Kubo formalism used above is limited to linear response, it is not immediately obvious how to evaluate the increase in energy or change in the Hamiltonian $H$. Simply replacing the operator $B$ above with $H$ will not work. However, for the present model it is possible to identify the dissipation directly with the work done.
 Work per unit time is force times velocity. During the perturbation period  it is not obvious or possible how to distinguish between reversible change of energy and irreversible
  dissipation of energy.  But this problem can be circumvented  by considering the total energy change due to a perturbation that lasts for a {\it finite time interval}. This is the
   method that we will  use  in the following.

Thus assume that the relative motion is finite in time, and  that
it starts at $t=0$  with maximum velocity $\bf v$ when the
position is ${\bf r=r}_0$. As $t \rightarrow \infty$ the motion is
required to die out. To accomplish this we introduce the
convergence factor $e^{-\eta t}$ ($\eta \rightarrow 0$) already
used in Eq.~(\ref{16}). In the interaction (\ref{2}), $t$ is thus
to be replaced with $q(t)=te^{-\eta t}$. As
mentioned we consider only the time interval $0 \leq t<\infty$, in
which the velocity decays exponentially,
\begin{equation}
{\bf v} \rightarrow {\bf v}_1(t)={\bf
v}\dot{q}(t)={\bf v}(1-\eta t)e^{-\eta t}. \label{20}
\end{equation}
For $\eta t >0$ , ${\bf v}_1(t)$ will now replace $\bf v$ in expression (\ref{19}) for the friction force. Altogether, the total energy dissipated will be
\begin{equation}
\Delta E_d=\int_{-\infty}^\infty {\bf v}_1(t)\cdot
{\bf F}_f\, \dot{q}(t)dt={\bf v \cdot F}_f\int_0^\infty
[\dot{q}(t)]^2dt=\frac{1}{4\eta}{\bf v\cdot F}_f, \label{21}
\end{equation}
where ${\bf F}_f$ is given by Eq.~(\ref{19}). Note that the
reversible part of the force ${\bf F}_r \propto
t\rightarrow q(t)$ as given by Eq.~(\ref{8}) will not contribute
to the dissipation since $\int_{0}^\infty
\dot{q}(t)q(t)dt=0$.

The above result for dissipation may be extended to more general
cases for which $q(t)$ shows an arbitrary
variation with time. Thus $q(t)$ need no longer
be linked to a varying position in space as shown in
Eq.~(\ref{1}). Under quite general conditions we may reinterpret
the situation as one for which $q(t)$ is
associated with a "position",
\begin{equation}
x(t)=q(t). \label{22}
\end{equation}
With this the "velocity" becomes
\begin{equation}
v(t)=\dot{q}(t). \label{23}
\end{equation}
As $x(t)$ is a scalar quantity the corresponding operator $B$ for the "force" will be
\begin{equation}
 B=-\frac{\partial}{\partial x}(-Aq(t))=A. \label{24}
\end{equation}
With relation (\ref{4}) the resulting "force" due to the perturbation becomes
\begin{equation}
F_f=\int_{-\infty}^t\phi_{AA}(t-t')q(t')dt',
\label{25}
\end{equation}
where
\begin{equation}
\phi_{AA}(t)=\frac{1}{i\hbar}{\rm Tr} \left\{ \rho [A, A(t)]\right\}. \label{26}
\end{equation}
With the associations made above  the total dissipated energy becomes
\begin{equation}
 \Delta E_d=\int_{-\infty}^\infty v(t)F_fdt=\int_{-\infty}^\infty\dot{q}(t)\left[\int_{-\infty}^t \phi_{AA}(t-t')q(t')dt'\right] dt. \label{27}
\end{equation}
The expression (\ref{21}) is consistent with this more general
result. In the integral (\ref{21}) a reversible part of the force
was taken out. This corresponds to writing
\begin{equation}
q(t')=q(t)-\dot{q}(t)(t-t')+... \label{28}
\end{equation}
The first term will not contribute to $\Delta E_d$, and we get
($q(t)=0$ for $t<0$)
\begin{equation}
\Delta E_d=-\int_0^\infty
\phi_{AA}(u)udu\int_{0}^\infty \left[ \dot{q}(t)\right]^2dt+...,
\label{29}
\end{equation}
which is the result (\ref{21}). This is seen by use of
Eqs.~(\ref{3}), (\ref{6}) and  (\ref{9}). Then with
$q(t)=t \rightarrow te^{-\eta t}$ one has $A=\bf
v\cdot B$ and thus $\phi_{AA}={\bf{v\cdot \phi}}_{BA}={\bf{ v\cdot
G}}\,\phi$, by which $ -\int_0^\infty \phi_{AA}\,udu={\bf {v \cdot
F}}_f$.

 The expression (\ref{27}) is thus the energy dissipation due to a perturbation  $-Aq(t)$ of the Hamiltonian. It is a general result, not necessarily limited to  the simple model given by Eq.~(\ref{1}).
Since we used an indirect argument implying the net work done on
an equivalent system, it would be of interest to verify this
result by an independent method. The recent approach of Barton
(personal communication) is in this context  of  interest, as he
makes use of quantum mechanical perturbation theory to the second
order in the interaction strength. If it could be shown that the
different approaches lead to the same results for force and
dissipation, it would be physically instructive.

\bigskip

{\bf Acknowledgment}

\bigskip

I. B. thanks Gabriel Barton for valuable correspondence.


\end{document}